\documentclass{elsart}
\usepackage{natbib}

\usepackage{graphics}
 \usepackage{graphicx}
 \usepackage{epsfig}
\usepackage{amssymb}

\begin{document}

\begin{frontmatter}


\title{A WENO algorithm for the growth of ionized regions at
   the reionization epoch}

\author[label1]{Jing-Mei Qiu},
\author[label1]{Chi-Wang Shu},
\author[label2]{Ji-Ren Liu},
\author[label2]{Li-Zhi Fang}

\address[label1]{Division of
Applied Mathematics, Brown University, Providence, RI 02912, USA}
\address[label2]{Department of Physics, University of Arizona,
Tucson, AZ 85721, USA}

\begin{abstract}

We investigate the volume growth of ionized regions around UV photon
sources with the WENO algorithm, which is an effective solver of
photon kinetics in the phase space described by the radiative transfer
equation. We show that the volume growth rate, either of isolated
ionized regions or of clustered regions in merging, generally consists
of three phases: fast or relativistic growth phase at the early
stage, slow growth phase at the later stage, and a transition phase
between the fast and slow phases. The growth rate can be
characterized by a time scale $t_c$ of the transition phase, which
is approximately proportional to $\dot{E}^{1/2}$, $\dot{E}$ being
the intensity of the ionizing source. The larger the time scale $t_c$,
the longer the photons to postpone their contribution to the
ionization. For strong sources, like $\dot{E} \geq 10^{56}$ erg
s$^{-1}$, $t_c$ can be as large as a few Myrs, which could even be
larger than the lifetime of the sources. Consequently, most photons
from these sources contribute to the reionization only when these
sources already ceased. We also show that the volume growth of ionized
regions around clustered sources with intensity $\dot{E}_i$ ($i=1, 2, \dots$)
would have the same behavior as a single source with intensity
$\dot{E}=\sum_i\dot{E}_i$, if all the distances between nearest
neighbor sources $i$ and $j$ are smaller than $c(t^i_c+t^j_c)$,
$t^i_c$ being the time scale $t_c$ of source $i$. Therefore, a tightly
clustered UV photon sources would lead to a slow growth of ionized
volume. This effect would be important for studying the redshift-dependence
of 21cm signals from the reionization epoch. We also developed, in
this paper, the method of using WENO scheme to solve radiative
transfer equation beyond one physical dimension. This method can 
be used for high dimensional problems in general as well.

\end{abstract}


\end{frontmatter}

\section{Introduction}

In the early stage of reionization of the universe, 
radiation from the first generation of stars ionizes neutron
hydrogen and helium to produce ionized bubbles around these stars.
The subsequent growing, overlapping and merging of these isolated
ionized patches lead to a full reionization of the universe. The
evolution of the reionization depends on the birth rate of the first
stars and the formation of ionized regions around these sources. The
growth of ionized HII volume is directly related to the formation of
21 cm emission and absorption regions at the reionization epoch
(e.g. Cen 2006; Alvarez et al. 2006; Chuzoy et al. 2006; Liu et al.
2007). The planning and ongoing projects of detecting redshifted 21
cm signals are trying to reconstruct the redshift evolution of the
reionized volume. Therefore, a detailed study on the merging of
ionized regions is necessary.

The growth of the ionized HII volume, $V$, is usually described by a
rate equation (e.g. Shapiro \& Giroux 1987;  Madau, Haardt \& Rees
1999; Mellema et al. 2006):
\begin{equation}
\label{eq1}
n(t)\frac{dV}{dt}=\dot{N} -\int_{V}
    n^2(t)\alpha_B C(t)dV,
\end{equation}
where $n=1.88\times 10^{-7}(\Omega_bh^2/0.022)(1+z)^3$ cm$^{-3}$ is
the mean number density of hydrogen at redshift $z$, $\dot{N}$ is
the emission rate of ionizing photons, $\alpha_B$ is the
recombination coefficient, and $C(t)$ is the volume-averaged
clumping factor of HII.

 From equation (1), $V(t)$ is linearly dependent on $\dot{N}$: $V(t)
\propto \dot{N}$. That is, the growth of the ionized volume is
proportional to the emission rate of ionizing photons $\dot{N}$,
regardless of whether the ionizing photons are produced from one
source with emission rate $\dot{N}$ or from $m$ sources with
emission rate $\dot{N}/m$. The linear relation between $V(t)$ and
$\dot{N}$ has been used in the simulations of the reionization (e.g.
Ciardi et al. 2003; Iliev et al. 2006). That is, the effects of the
photon kinetics in the phase space are completely ignored.

A basic assumption of equation (1) is that photons emitted from the
sources will immediately join the action with the atom, regardless
of the photon propagation between the source and the atom. This
assumption is reasonable if the time scale of the photon kinetics is
much smaller than that of the problem considered. Unfortunately, this
is not always correct for problems at the epoch of reionization
(Shapiro et al. 2006; Qiu et al. 2007). In this paper, we will show
that the finite speed of light will lead to a substantial change of the
growth rate of the ionized volume. The ionized volume growth of
one source with emission rate $\dot{N}$ can be significantly
different from that of multiple sources with the total emission rate
equal to $\dot{N}$. The growth rate of the ionized volume depends not
only on the total emission rate of ionizing photons, $\dot{N}$, but
also on the distribution and clustering of the sources.

Many numerical solvers for the radiative transfer equation have been
proposed (Razoumov \& Scott 1999; Abel et al. 1999; Ciardi et al.
2001, Gnedin \& Abel 2001, Sokasian et al. 2001, Nakamoto et al.
2001; Razoumov et al. 2002, Cen 2002, Maselli et al. 2003, Shapiro
et al., 2004; Rijkhorst et al. 2006; Mellema et al. 2006; Susa 2006,
Whalen \& Norman 2006). We use the WENO scheme to be the solver for
the photon kinetics in the phase space. The WENO algorithm has been
proved to have high order accuracy and good convergence in capturing
discontinuities and complicated structures in fluid as well as to be
significantly superior over piecewise smooth solutions containing
discontinuities (Shu 2003). We have showed that the WENO algorithm
is effective for solving radiative transfer problem in
one-dimensional physical space and one-dimensional frequency space
(Qiu et al. 2006, 2007). It revealed that the time-dependent
solution of the radiative transfer equations is essential for the
formation and evolution of the ionized and heated regions around UV
ionizing sources. In this paper, we will develop the WENO algorithm
of the radiative transfer equations beyond one-dimension.

The paper is organized as follows. Section 2 presents the
source-intensity dependence of the growth rate of ionized regions
around isolated point sources. Section 3 studies the merging of two
ionized regions and its effect on the growth of the ionized volume.
Discussions and conclusions are given in Section 4. The details of
the WENO numerical scheme are listed in Appendix.

\section{Isolated point source}

The patchy structures of the HII region in the early universe is
very complex. However, in the early stage, many HII regions are
isolated and even spherical around UV photon sources. Later, these
spherical regions merge and yield complicated structures. As a
preparation, we summarize in this section, the features of the
growth of the isolated ionized regions, calculated
with the WENO algorithm (Qiu et al. 2006, 2007, Liu et al. 2007).
To make the paper self-contained, the corresponding equations and
parameters are given in Appendix A.

\subsection{The growth of the ionized volume}

In deriving eq.(\ref{eq1}), it is assumed that the ionization in
region $V$ is perfect, i.e., the fraction of the neutral hydrogen,
$f_{\rm HI}\equiv n_{\rm HI}/n_{\rm H}$, is zero within the region,
and 1 outside. Actually, the ionization cannot be complete,
and the ionization front (I-front) cannot be defined by a sharp boundary
dividing the completely ionized and completely neutral regions. We
will define the ionized region to be the place in which
$f_{\rm HI}<90\%$.

Consider a point source emitting photons of energy
$\dot{E}(\nu)d\nu$ per unit time within the frequency ranges from
$\nu$ to $\nu+d\nu$. The energy spectrum of photons is assumed to be
a power law, $\dot{E}(\nu)=\dot{E}_0(\nu_0/\nu)^{\alpha}$ with
$\alpha = 2$, $\nu_0$ is the ionization energy. Assuming the
hydrogen gas around the source is uniform, the volume growth of the
ionized region, $V(t)$, at redshift $1+z=10$, is shown in Figure 1,
in which the intensities of UV photons are taken to be
$\dot{E}=\int_{\nu_0}^{\infty} E(\nu)d\nu=5.8\times 10^{39}$,
$10^{41}$, $10^{43}$ and $10^{45}$ erg s$^{-1}$, or
$\dot{N}=1.34\times10^{50}$, $10^{52}$, $10^{54}$ and $10^{56}$
s$^{-1}$.

In Figure 1, we use Myrs and Mpc to be the units of time $t$ and
length $r$. In this paper, we also sometimes use dimensionless time
and length defined as $t' = c n \sigma_0 t$ and  $r' = n \sigma_0
r$, where $\sigma_0$ is the ionization cross section. Therefore,
$t=0.89 (1+z)^{-3} t'$ Myrs and $r=0.27(1+z)^{-3}r'$ Mpc. $t'$ and
$r'$ actually are in the units of mean free flight time and mean free
path of ionizing photons. The dimensionless variables are convenient
for numerical works (see Appendix C).

\begin{figure}
\centering
\includegraphics[width=10cm]{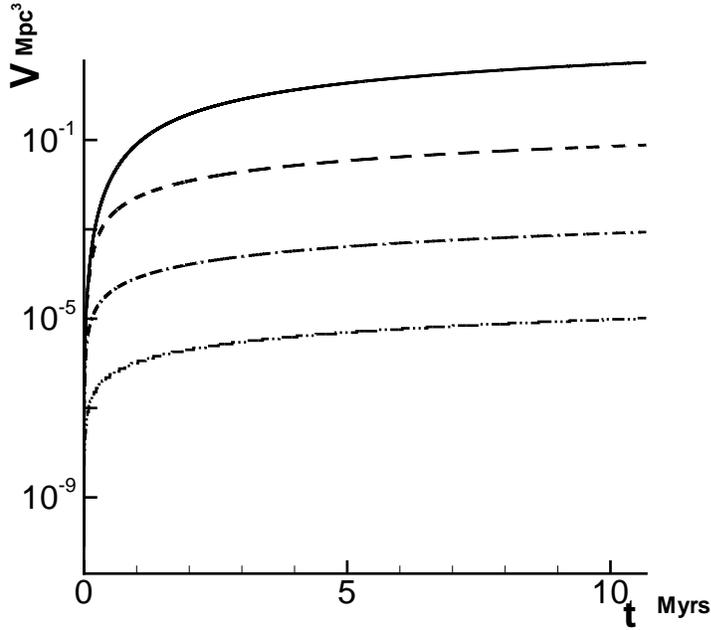}
\caption{Ionized volume $V(t)$ vs. time $t$. The source intensities
are taken to be $\dot{E}=5.8 \times [10^{39}$ (dash dot dot),
10$^{41}$ (dash dot), 10$^{43}$ (dash), and 10$^{45}$ (solid line)]
erg sec$^{-1}$, and $\dot{N}$ to be 1.34$\times [10^{50}$,
10$^{52}$, 10$^{54}$, and 10$^{56}$] sec$^{-1}$, respectively.
Redshift is taken to be $1+z=10$.}
\end{figure}

Figure 1 shows a common feature for all sources that the growth of
the ionized volume undergoes three phases: when $t$ is small, the
growth is very fast, when $t$ is large the growth is very slow, and
a transition phase between the fast and slow phases. The details of
the growth are different for different sources. For a weak source
with $\dot{E}=5.8\times 10^{39}$ erg s$^{-1}$, the ionized volume at
the end of the fast growth phase is already comparable with that at
later time. However, for a strong source with $\dot{E}=5.8\times
10^{45}$ erg s$^{-1}$ the ionized volume at the end of the fast
growth phase is much smaller than that at later time.  These
features can be seen more clearly in Figure 2, which plots $d\ln
V(t)/d\ln t$ vs. $\ln t$ for the same solutions in Figure 1. $d\ln
V(t)/d\ln t$ is the index $a$ of the power-law relation $V \sim
t^a$. Figure 2 shows once again the three phases of the power-law
index. When $t$ is small, $d\ln V(t)/d\ln t \simeq 3$, or
$V(t)\propto t^3$, and the radius of the ionized spheres, or the
I-front, $r_{\rm I}=(3V/4\pi)^{1/3} \simeq ct$. Therefore, this
actually is the fast or relativistic phase. In the transition phase,
$d\ln V(t)/d\ln t$ decreases from 3 to about 1. Finally, $d\ln
V(t)/d\ln t$ approaches to $\sim$ 1, or $r_{\rm I}\propto t^{1/3}$;
this is the slow or non-relativistic phase.

\begin{figure}
\centering
\includegraphics[width=9cm]{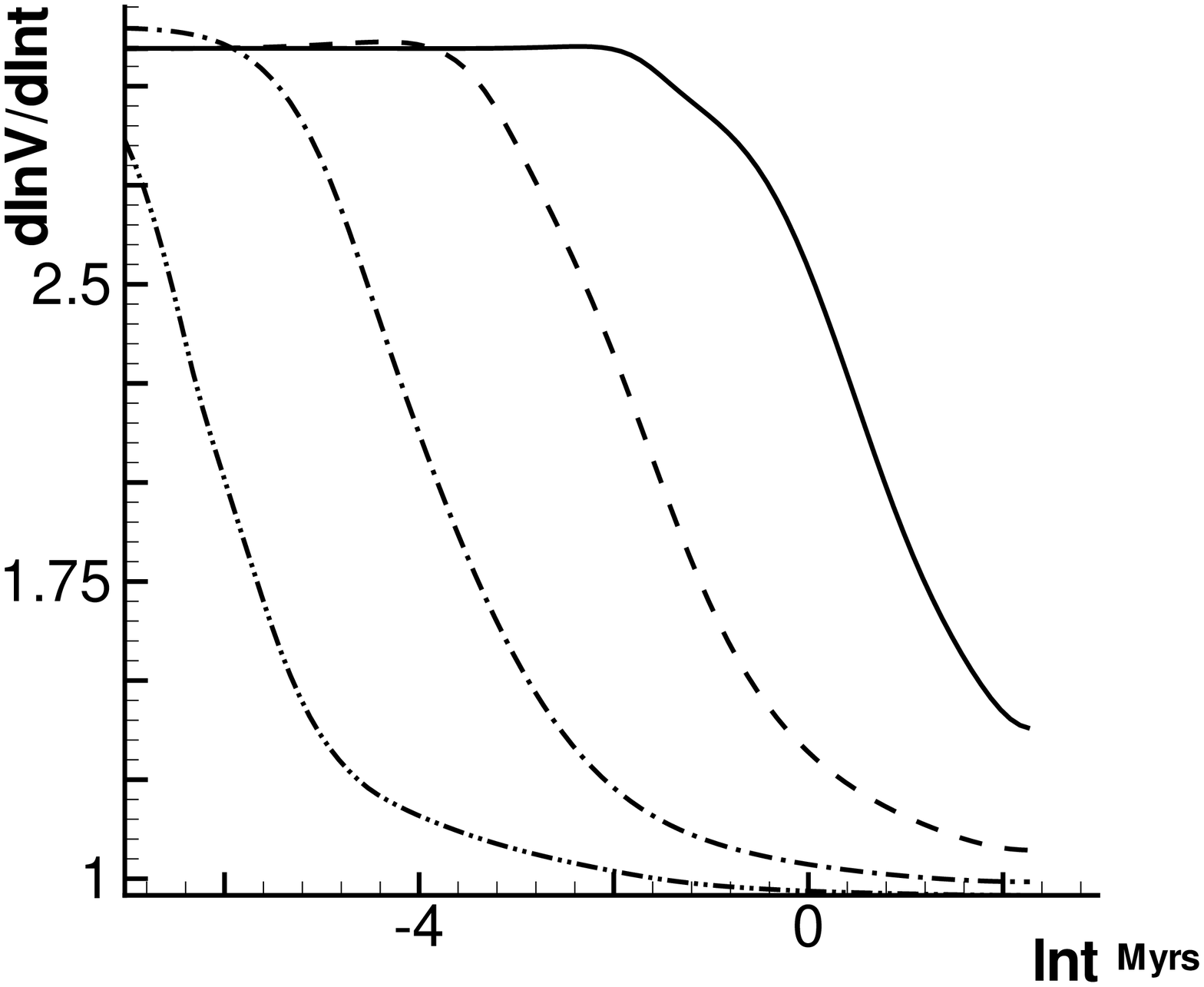}
\caption{$d\ln V(t)/d\ln t$ vs. $\ln t({\rm Myrs})$ . The source
intensities are taken to be $\dot{E}=5.8\times [10^{39}$ (dash dot
dot), 10$^{41}$ (dash dot), 10$^{43}$ (dash), and 10$^{45}$ (solid
line)] erg sec$^{-1}$.}
\end{figure}

\subsection{$\dot{E}$-dependence of the ionized region growth}

We define a time scale, $t_c$, by $d\ln V(t)/d\ln t|_{t=t_c}=2.5$,
which characterizes the transition time from the fast phase to the
slow phase. The dependence of the time scale $t_c$ on $\dot{E}$ is plotted in
Figure 3, which approximately follows  $t_c \propto (\dot{E})^{1/2}$.
Therefore, the transition time of the growth of the ionized volume
is non-linearly dependent
on the emission rate of the ionizing photon $\dot{E}$, which indicates
that the growth of $V(t)$ also depends non-linearly on $\dot{E}$.

\begin{figure}
\centering
\includegraphics[width=9cm]{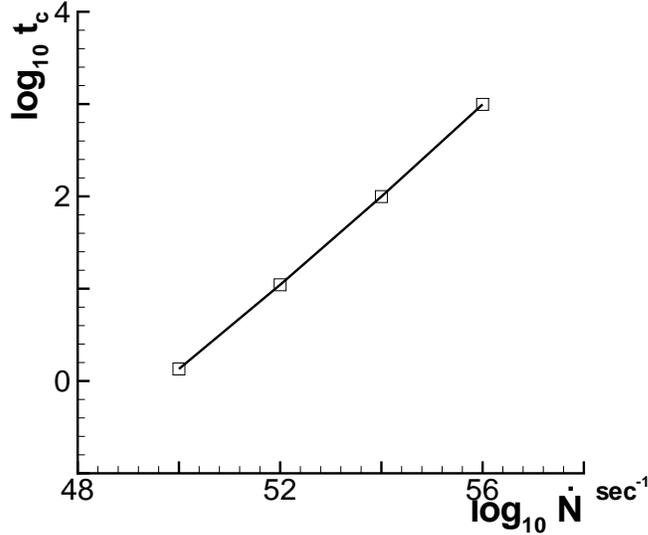}
\caption{ $\log_{10}(t_{c})$ vs. $\log_{10}{\dot{N}}({\rm
sec}^{-1})$. Other parameters are the same as those in Figures 1 and
2. $t_{c}$ is in the unit of mean free flight time of ionized
photons.}
\end{figure}

The non-linear dependence of $V(t)$ on $\dot{E}$ can be more prominently
demonstrated by Figure 4, in which we plot the growth of the total ionized
volume $V_{\rm total}(t)$ around one source with $\dot{E}=5.8\times 10^{45}$ erg
s$^{-1}$, and those given by $m$ sources with intensity $5.8\times 10^{45}/m$
erg s$^{-1}$, with $m=10^2$, $10^4$ and $10^6$. Here we assume that
the ionized regions for different sources do not overlap.
Figure 4 shows that the growth rate of the ionized volume depends
substantially on the number of sources, in spite of the
fact that in all cases the total photon emission rate, $\dot{N}$,
are the same. For a single strong source of
$\dot{E}=5.8\times 10^{45}$ erg s$^{-1}$, $V_{\rm total}(t)$ at
$t \simeq 1$ Myrs is only about 0.16, 0.10 and 0.08 of that from $m$
sources with intensities $\dot{E}=5.8\times 10^{45}/m$, erg s$^{-1}$
and $m=10^2$, $10^4$ and $10^6$.

Therefore, the larger the $t_c$, the less effective the ionization
with the same rate $\dot{N}$. This is simply due to the retardation
of photon propagation. In the period of $t<t_c$, the I-front
propagates with about the speed of light $c$, and therefore,
most photons emitted later will delay their contribution to the
reionization by a time $t_c$. The longer the $t_c$, the longer the
delay. When $t$ is larger than $t_c$, the speed of the I-front is
slowing down and the later emitted photons start to join the
reionization.

It is interesting to compare Figures 2 and 4. Figure 2 shows that
the ionized volume growth velocity $d\ln V(t)/d\ln t$ approaches to
1 at $t\simeq 5$ Myrs for all sources with $\dot{E}\leq 5.8\times
10^{45}$ erg s$^{-1}$. However, Figure 4 shows that the total
ionized volume $V_{\rm total,1\times 45}(t)$ of one source
$\dot{E}=5.8\times 10^{45}$ erg s$^{-1}$ at 5 Myrs is still much
less than the total ionized volume $V_{\rm total, m\times 45/m}(t)$ of
$m$ ($>1$) sources with $\dot{E}=5.8\times 10^{45}/m$ erg s$^{-1}$.
That is, the total ionized volume of one source $\dot{E}=5.8\times
10^{45}$ does not catch up with the total ionized volume of $m$ sources with
$\dot{E}=5.8\times 10^{45}/m$ erg s$^{-1}$ even when they have about
the same growth velocity $d\ln V(t)/d\ln t\simeq 1$. This is because
$dV_{\rm total, m\times 39/m}/dt$ is larger than or equal to
$dV_{\rm total, 1 \times 45}/dt$ for all time $0< t <t_{rec}$, we
have $V_{\rm total, m\times 45/m}(t) -V_{\rm total, 1 \times
45}(t)=\int_0^t [dV_{\rm total, m\times 45/m}/dt -dV_{\rm total, 1
\times 45}/dt]dt
>0$ in the period  $t < t_{rec}$, i.e. before
the ionized regions approach their Stromgren sphere. At $1+z =10$,
$t_{rec}\simeq 8.6\times 10^8$ yrs, which is comparable with $1/H$,
and therefore part of the UV photons from strong sources may
cause ionization when the sources have already ceased. Comparing
with weak sources, the reionization of strong sources such as
$\dot{E} \geq 10^{45}$ erg s$^{-1}$ is less effective at the period
$t<t_{rec}$.

\begin{figure}
\centering
\includegraphics[width=9cm]{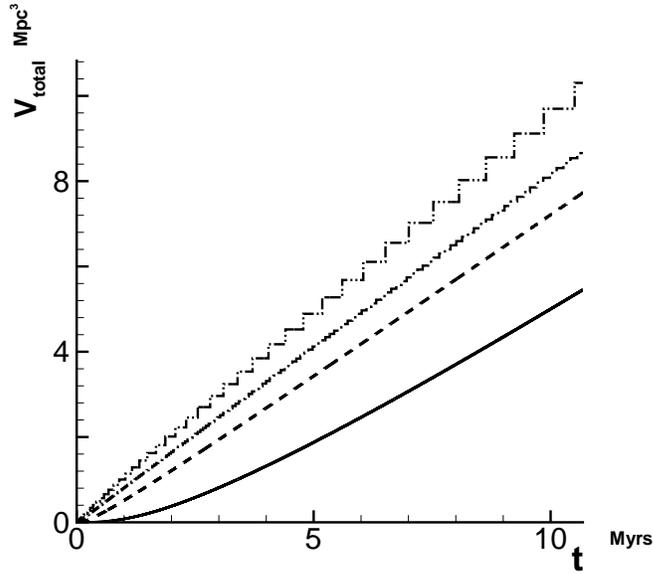}
\caption{Evolution of $V_{\rm total}(t)$. The source intensities are
taken to be $\dot{E}=5.8\times [10^{39}$ (dash dot dot),
10$^{41}$(dash dot), 10$^{43}$ (dash), and 10$^{45}$ (solid line)]
erg sec$^{-1}$.}
\end{figure}

\subsection{Inhomogeneous distribution of gas}

\begin{figure}
\centering
\includegraphics[width=9cm]{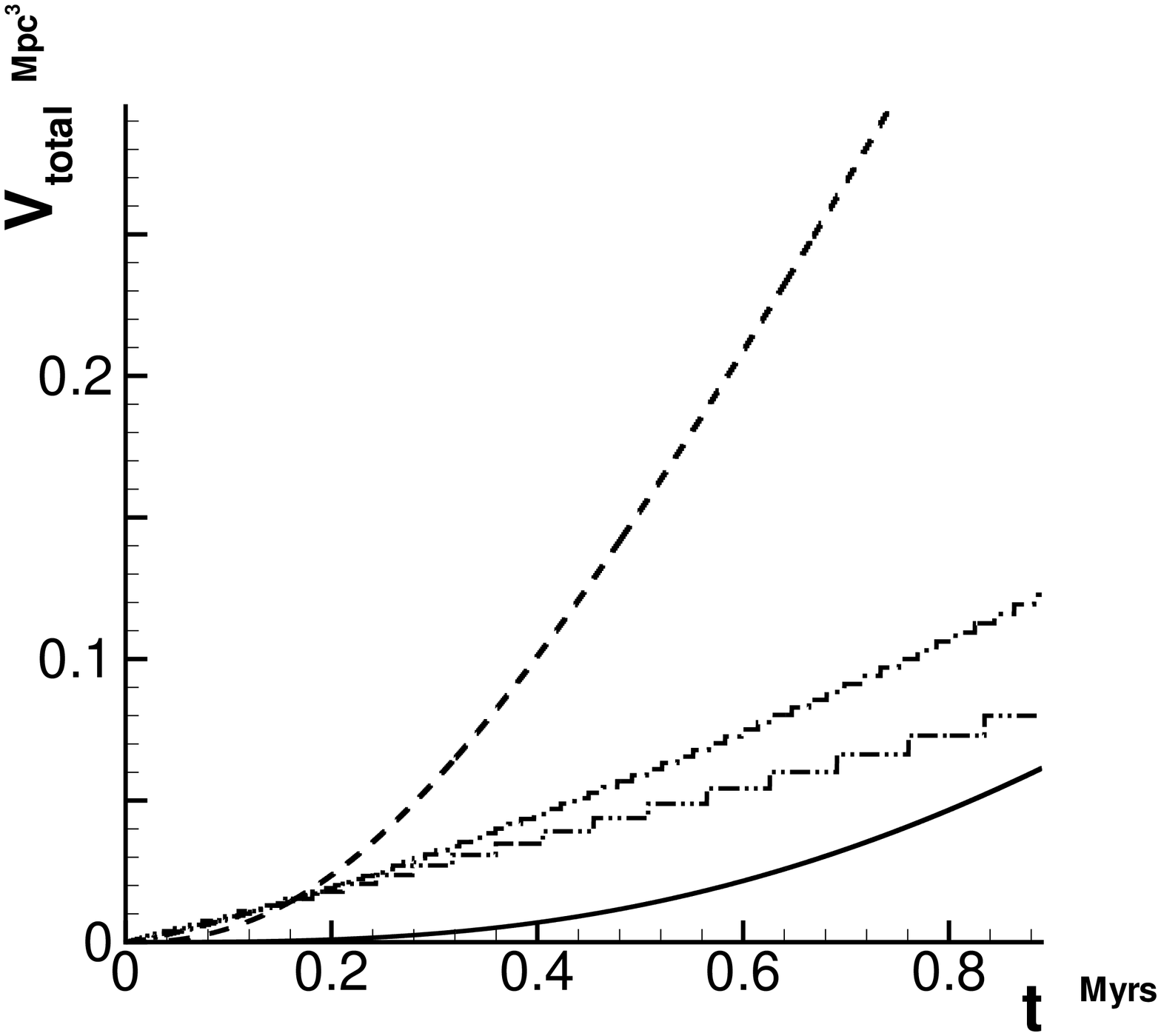}
\caption{The evolution of $V_{\rm total}(t)$ for the number density
distribution given by eq.(\ref{eq2}). The source intensities are
taken to be $\dot{E}=5.8\times [10^{39}$ (dash dot dot), 10$^{41}$
(dash dot), 10$^{43}$ (dash), and 10$^{45}$ (solid line)] erg
sec$^{-1}$. }
\end{figure}

Generally, the mass density of gas is high near the source. To study
its effect, we assume the density distribution $n(r)$ is given by
\begin{equation}
\label{eq2}
\frac{n(r)}{n_0} =  1+ \frac{n_c}{n_0}e^{-r/R},
\end{equation}
where $R$ is the size of the high density region, and $n_c/n_0$ is
the density increase in the center of the sources. As an
example, we choose $(n_c/n_0)=10$ and $R=50$ in the unit of the mean
free path of the ionization photons. The growth of the ionized
volume from different intensity of sources is plotted in Figure 5,
which shows the similar behavior as that in Figure 4.
Quantitatively, the high density core of $n_r/n_0=10$ and $R=50$ plays
a similar role as a sphere with size 500 in the unit of mean free path.
We have also calculated the evolution using other inhomogeneous
density models and have obtained results similar to those in Figure 5.

\section{Clustered point sources}

\subsection{Time scale with the merged ionized regions}

The UV photon sources of the first generation of stars may not be very
strong, however, they are most likely clustered. The merging of
ionized regions around a single source will lead to complicated
configuration of the ionized regions of clustered sources. However,
in terms of the growth rate of the ionized volume, the problem is
simplified. The merging process of clustered ionized regions can
approximately be decomposed into a set of two-region merging. It is
similar to the identification of clusters by the friend-of-friend
method. Therefore, we may reveal some common features of the merging
effect on the ionized volume growth rate by a detailed study of the
merging of two ionized regions.

\begin{figure}
\centering
\includegraphics[width=7cm]{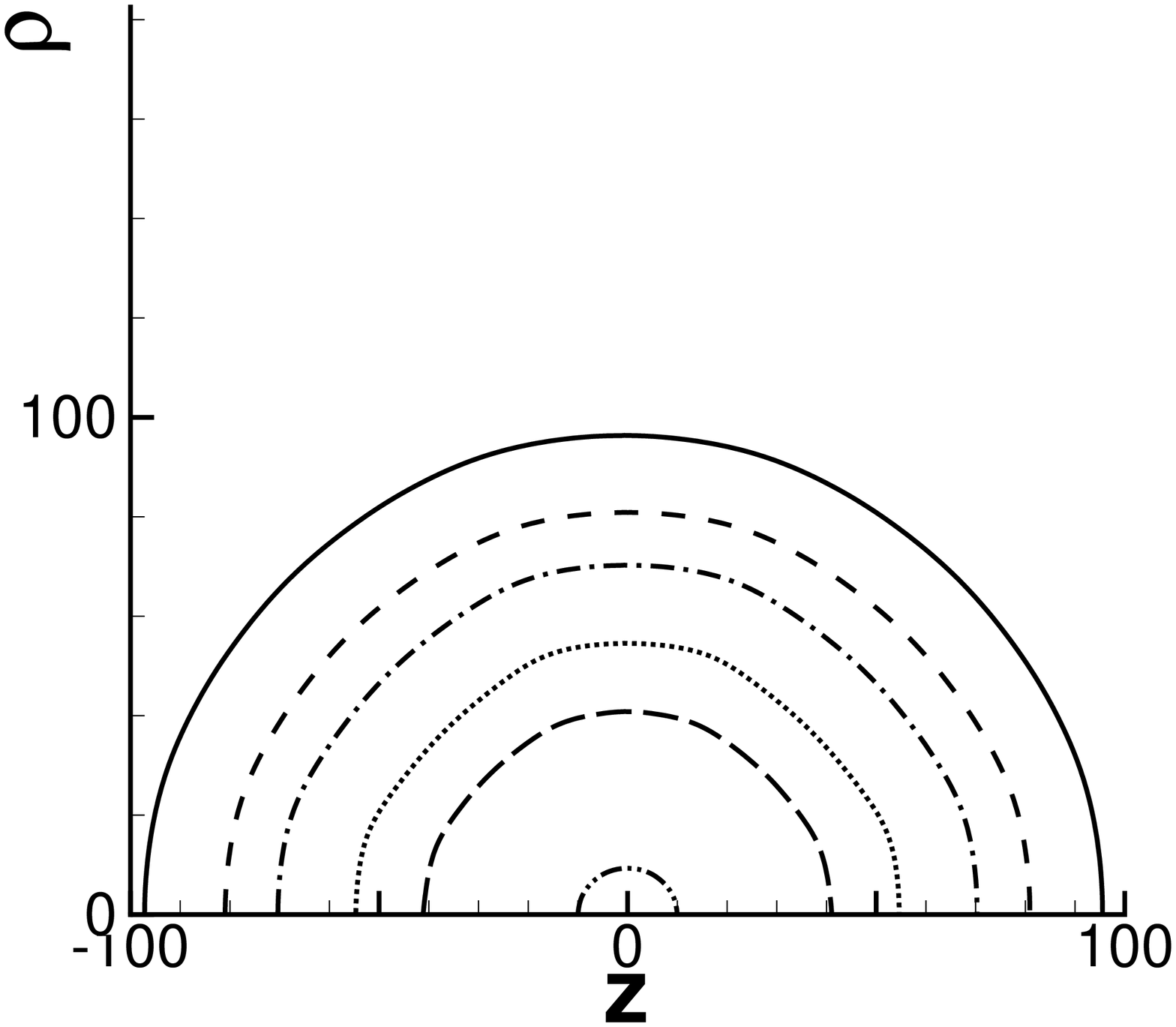}
\includegraphics[width=7cm]{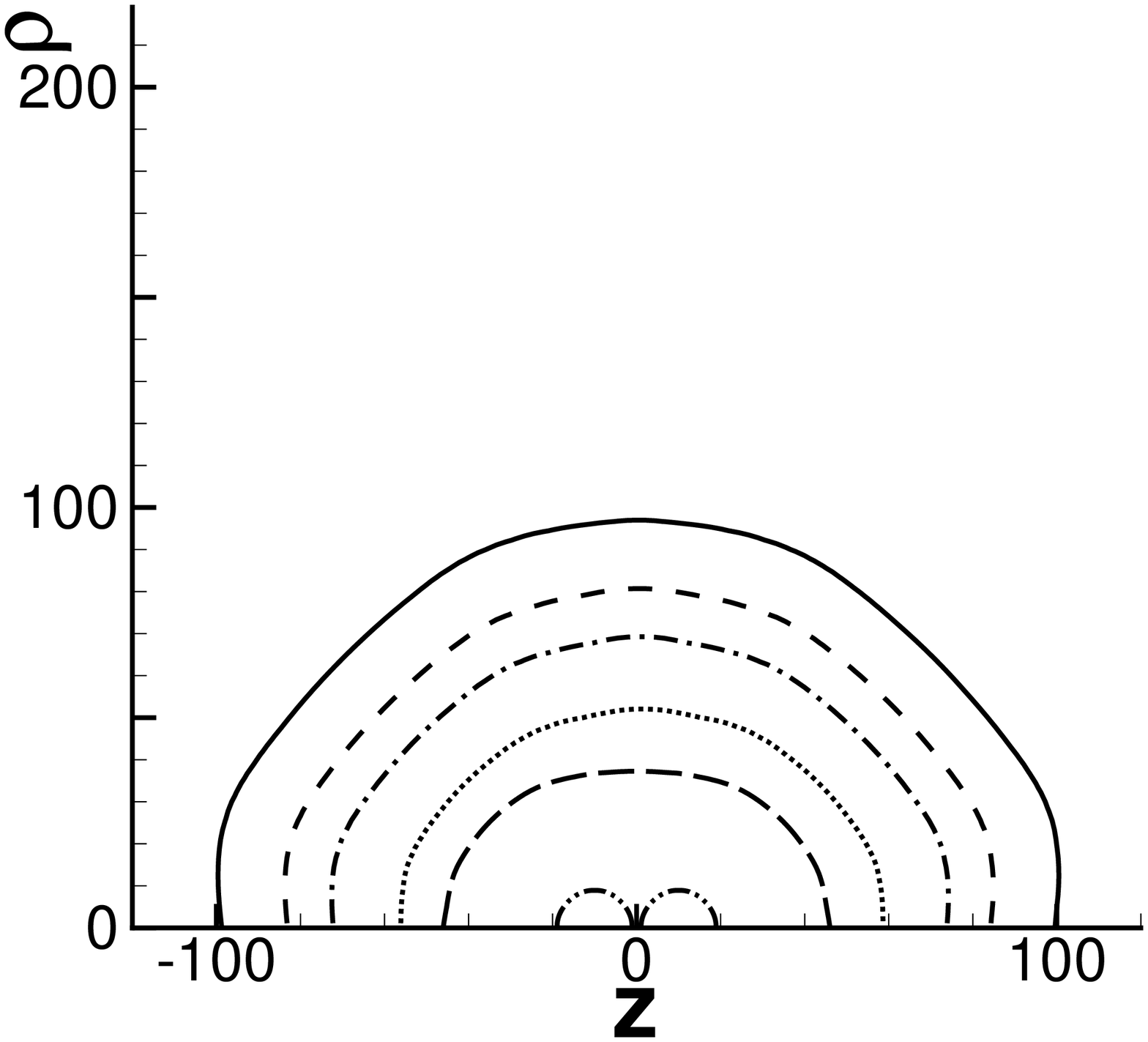}
\includegraphics[width=7cm]{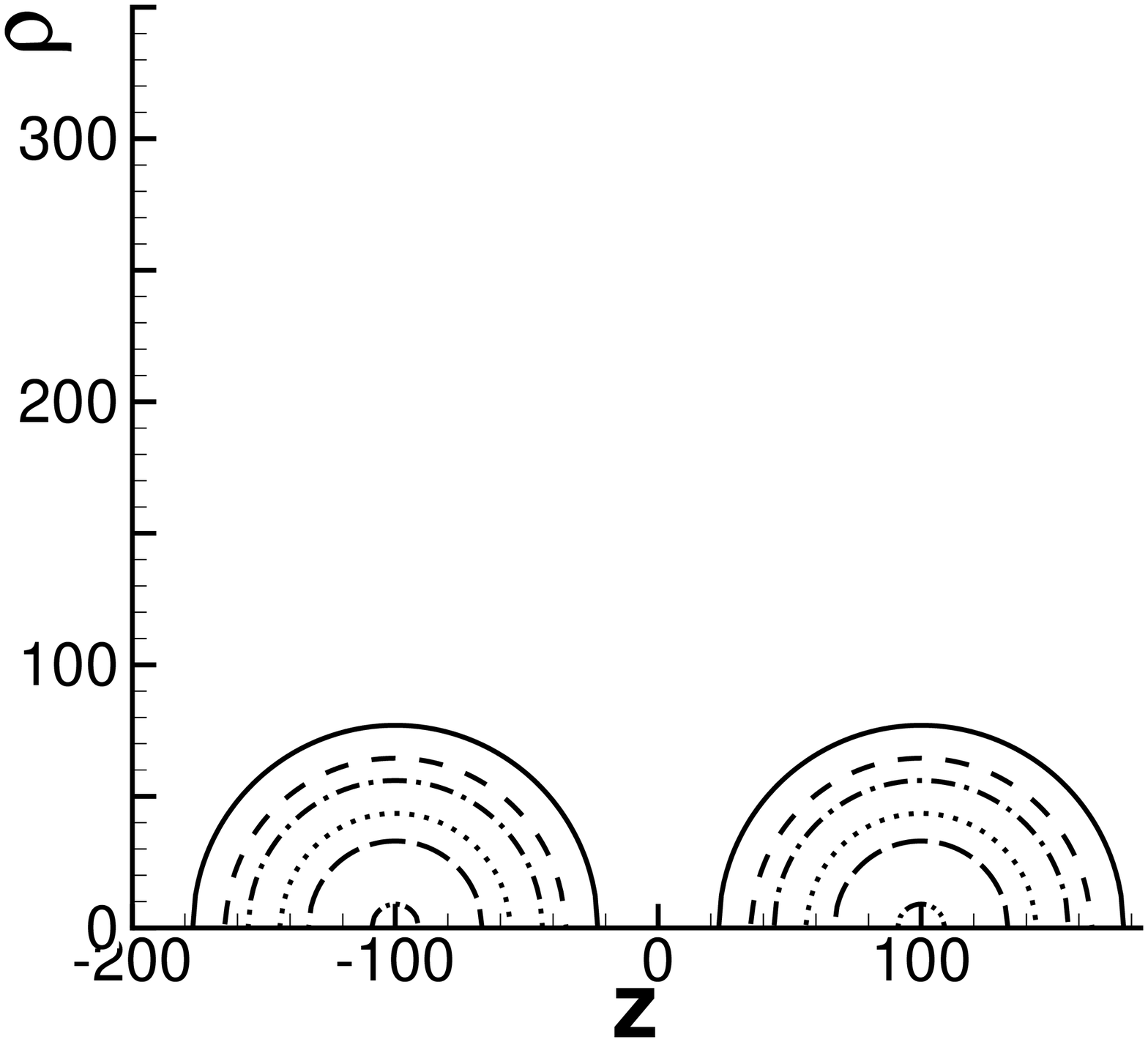}
\caption{ The evolution of the ionized region ($f_{\rm HI}(\rho, z,
t)<0.90$) of two UV photon sources with intensity $\dot{E}=0.5
\times 5.8\times 10^{41}$ erg s$^{-1}$. The contours from small to
large correspond, respectively, to the time $t'$ = 10 (dash dot
dot), 100 (long dash), 200 (dot), 400 (dash dot), 600 (dash), 1000
(solid) in the unit of the mean free flight time. $\rho$ and z are
dimensionless, i.e. in the unit of mean free path of ionizing
photons. The distance between the two sources is $a=1$ (top), 10
(middle) and 100 (bottom) also in the unit of mean free path of
ionizing photons. }
\end{figure}

Let us consider two sources located, respectively, at $(x,y,z)=(0,0,
a)$ and $(0, 0, - a)$. We calculate the time-dependence of the
profile of the ionized regions around the two sources. The evolution
of the profiles of the ionized region in the $\rho-z$ plane
($\rho=\sqrt{x^2+y^2}$) is shown in Figure 6, where the variables
$\rho$ and $z$ are dimensionless as defined in \S 2.1. The ionized
region is still defined by the region in which $f_{\rm HI}(\rho, z,
t)<0.9$. In Figure 6, the intensities of the two sources are taken
to be $\dot{E}=0.5\times5.8\times 10^{41}$ erg s$^{-1}$, and $a=1$,
10 and 100 in the unit of mean free path of the ionizing photons.
For each case, the profiles are at the time $t'= 10, 100, 200, 400,
600, 1000$ in the unit of the mean free flight time.

\begin{figure}
\centering
\includegraphics[width=7cm]{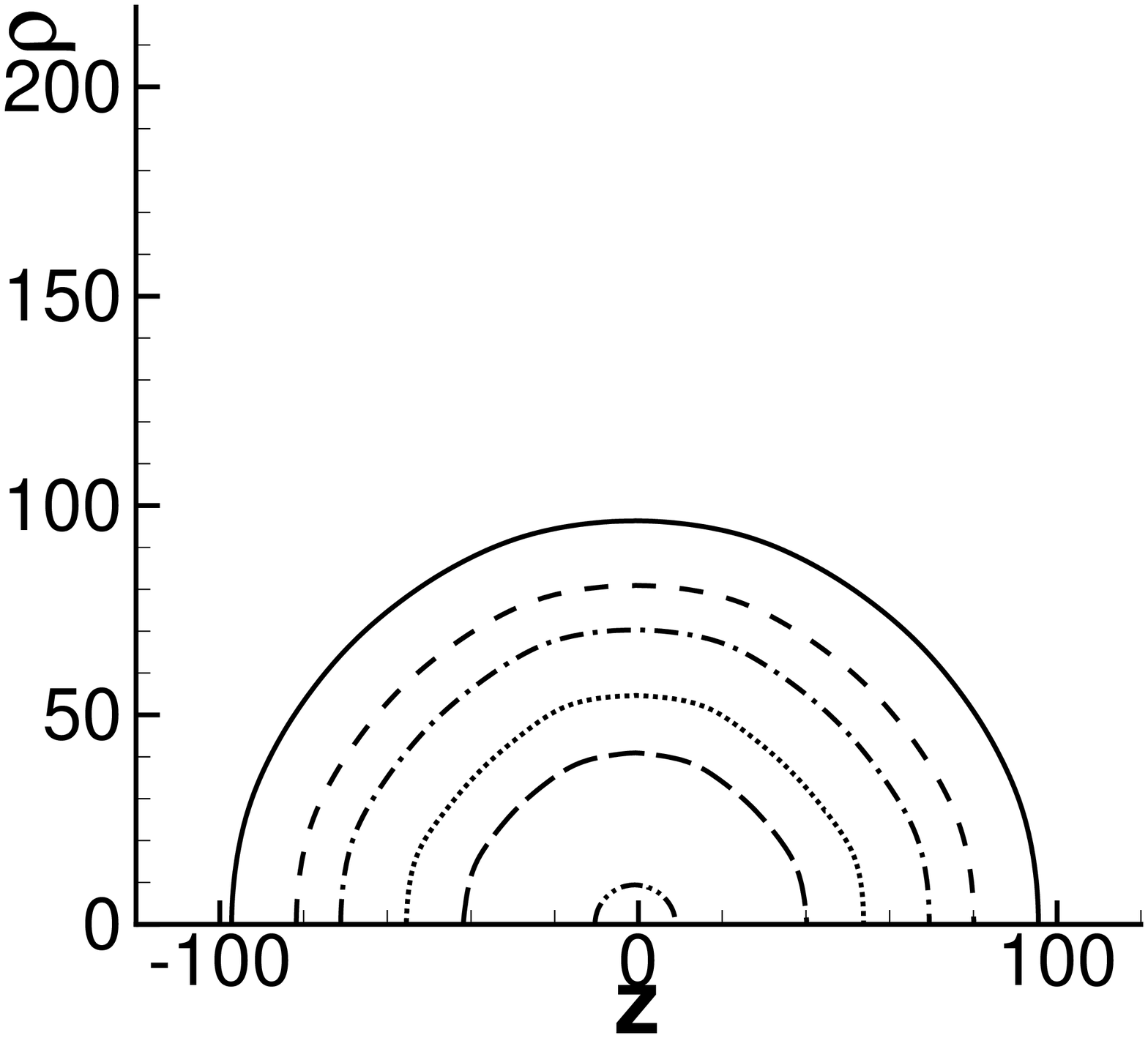}
\includegraphics[width=7cm]{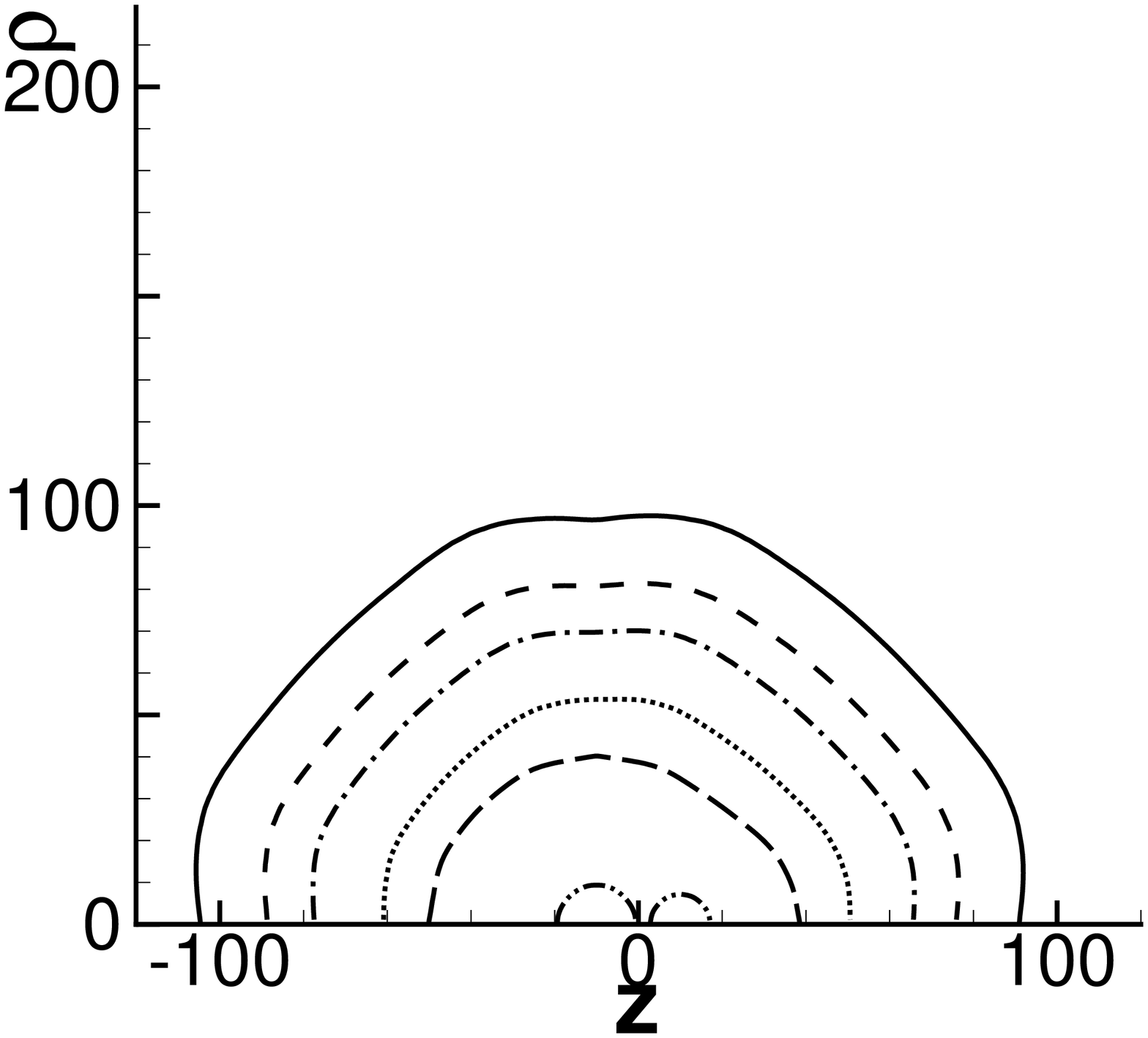}
\includegraphics[width=7cm]{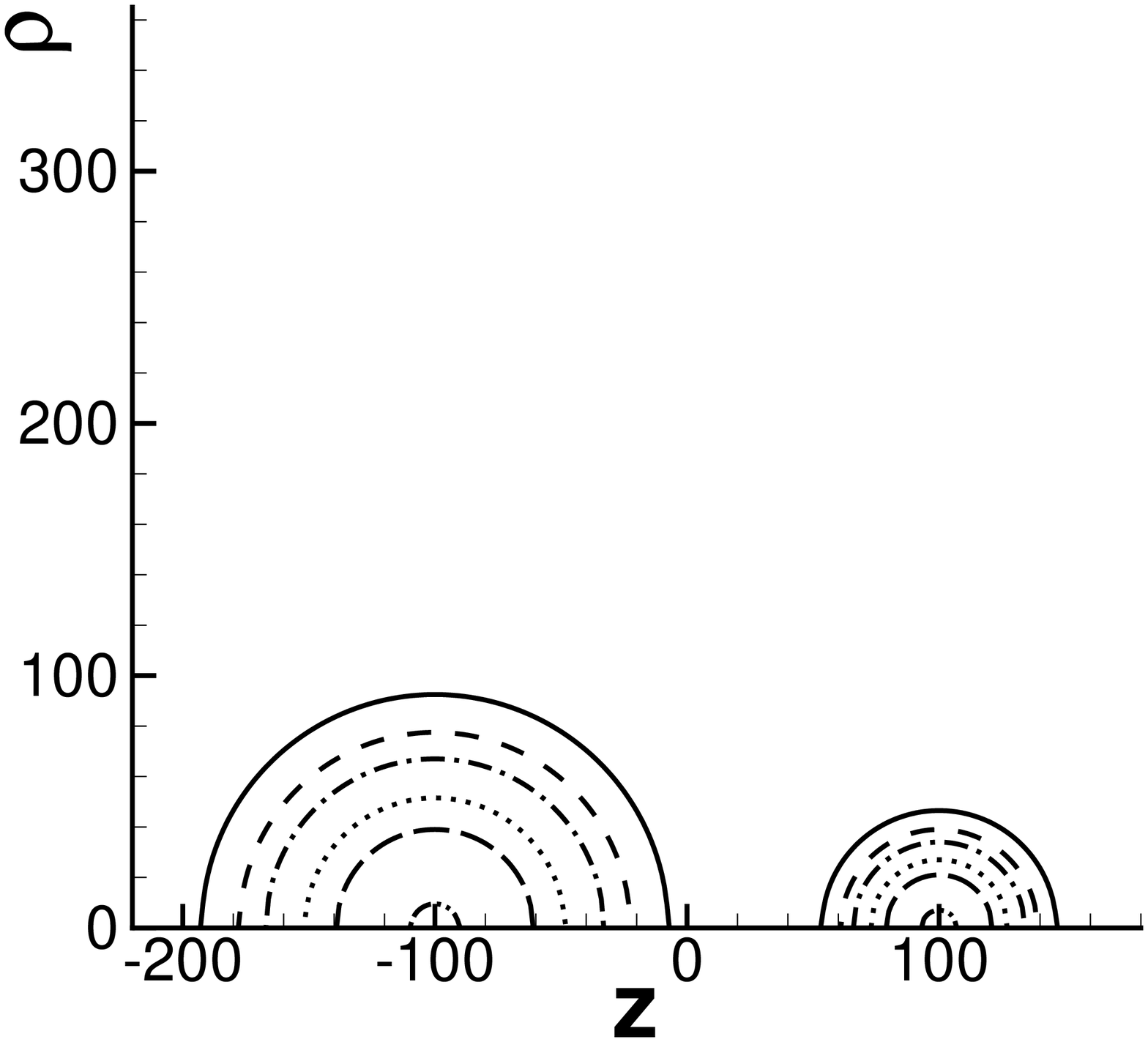}
\caption{
The evolution of the ionized region ($f_{\rm HI}(\rho, z, t)<0.90$)
of two UV photon sources with intensities $\dot{E}=0.9 \times 5.8
\times 10^{41}$ and $0.1\times5.8\times 10^{41}$ erg s$^{-1}$.
The contours from small to large sizes correspond, respectively,
to the time $t'$ = 10, 100, 200, 400, 600, 1000 in the unit of the
mean free flight time. $\rho$ and z are in units of mean free path
of ionizing photons. The distance between the two sources is $a=1$
(top), 10 (middle) and 100 (bottom)
in the unit of mean free path of ionizing photons.
}
\end{figure}

 From Figure 6, one can see first that the configuration of the
ionized regions is very different from the ionized sphere of a point
source. In the case of $a=1$, the ionized regions have already
merged when $t' < 10$, and the profile of the merged ionized region
is like a sphere. For $a=10$, there are two spheres around the two
sources when $t'<10$, and they merge at $t'<100$. The profile of the
merged ionized region is no longer spherical. For $a=100$, however,
no merging occurs even at the time $t' \simeq 1000$. This is simply
because the time scale $t_c$ of sources with $\dot{E}=0.5\times5.8
\times 10^{41}$ erg s$^{-1}$ is $\simeq 30$. If the distance between
the two sources is less than $ct_c$, the merging is realized in the
fast phase. On the other hand, for $a=100$, which is larger than
$ct_c$, the merging time will be much larger than $a/c$.

In Figure 7, we show the evolution of the ionized profiles
for two sources located, respectively, at $(x,y,z)=(0,0, a)$ and
$(0, 0, - a)$ with intensities $\dot{E}=0.9\times 5.8\times
10^{41}$ and  $\dot{E}=0.1\times 5.8\times 10^{41}$ erg s$^{-1}$.
Similar to Figure 6, in the case of $a=1$, the ionized regions have
merged when $t < 10$, and the profile of the merged ionized region
is like a sphere. For $a=10$, there are two spheres around the two
sources when $t<10$, and they merge at $t<100$.  There is no
merging for the case of $a=100$ even when the time is as large as $t
\simeq 1000$. Therefore, the basic feature of Figure 7 is the same
as that in Figure 6: for two sources with the transition time $t_{1c}$ and
$t_{2c}$, if their distance $2a$ is less than $c[t_{1c}+t_{2c}]$,
the merging occurs quickly, while it will be very slow if
$2a>c[t_{1c}+t_{2c}]$.

 From the middle panel of Figure 7, it is interesting to see that the
two ionized spheres have about the same size, although the
intensities of the two sources are different by a factor of about
10. This is because in the fast phase, the growth of the ionized sphere
radius is given by the speed of light, regardless of the intensity
of the sources.

\subsection{Growth of the ionized regions of two sources}

For the two source case, the configuration of the ionized regions
generally is very different from the spherical ionized region of a
point source. What we want to show in this section is, however, that
the growth of the total ionized volume, $V(t)$, of two sources with
intensities $\dot{E}_1$ and $\dot{E}_2$ is the same as that of a
single source of $\dot{E}=\dot{E}_1+\dot{E}_2$ if the distance
between the two sources $2a$ is less than $c(t_{1c}+t_{2c})$, with
$t_{1c}$ and $t_{2c}$ being the transition time scales of the two
sources, respectively. If $2a$ is larger than $c(t_{1c}+t_{2c})$,
the effect of the merging is small, and the growth of the total
ionized volume of the two sources can basically be treated as two
isolated ionized regions, i.e. its growth rate should be faster than
that of the single source of $\dot{E}=\dot{E}_1+\dot{E}_2$.

\begin{figure}
\centering
\includegraphics[width=9cm]{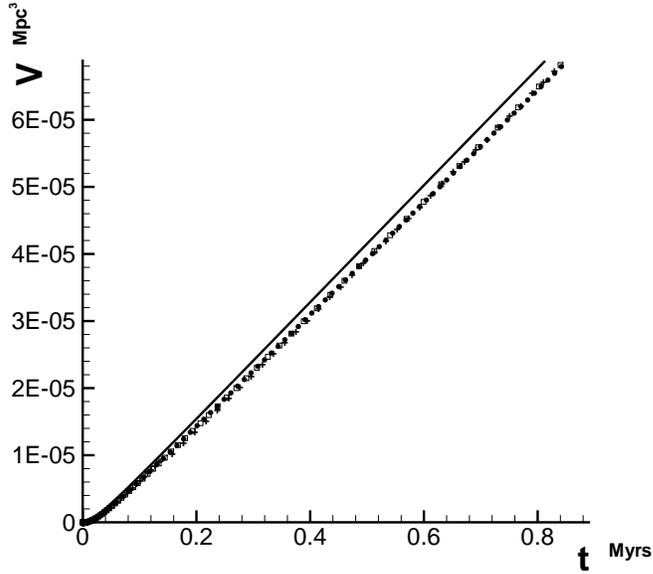}
\caption{Ionized volume $V(t)$ vs. time $t$ of two sources with the
same parameters as those in Figure 6. $a=100$, 10 and 1 are shown,
respectively, by solid line, filled circles and crosses. The unfilled
square symbols are for the single point source with
$\dot{E}=\dot{E}_1+\dot{E}_2$.}
\end{figure}

Figure 8 plots the evolution of $V(t)$ of two sources with
parameters $\dot{E}_1$ and $\dot{E}_2$ as those used in Figure 6 of
the last section. The $V(t)$ of a single source with intensity
$\dot{E}=\dot{E}_1+\dot{E}_2$ is also given in Figure 8. From Figure
3, we know that for $a=1$ and 10, $2a$ is less than
$c(t_{1c}+t_{2c})$, while for $a=100$, $2a$ is larger than
$c(t_{1c}+t_{2c})$. Figure 8 indeed  shows clearly that the growth
of the total ionized volume of $a=100$ is faster than that of $a=1$ and 10,
although the source intensity of $a=100$ is the same as that of $a=1$ and
10. The growth of the ionized volume of the cases $a=1$ and 10 are the
same. They are also exactly the same as the $V(t)$ of a single source
with intensity $\dot{E}=\dot{E}_1+\dot{E}_2$, though the
configuration of the ionized regions of two sources is very
different from that of a single point source.

\begin{figure}
\centering
\includegraphics[width=9cm]{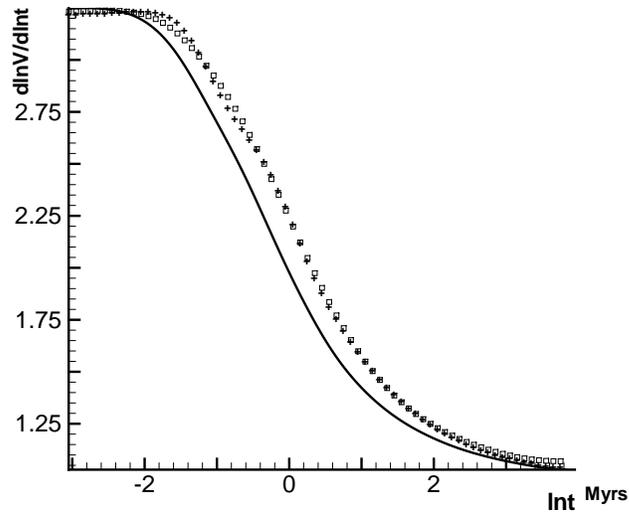}
\caption{$d\ln V(t)/d\ln t$ vs. $\ln t$ (variable $t$ is
dimensionless) of two sources with intensity $\dot{E}=(1/2)5.8\times
10^{41}$ erg s$^{-1}$ at $a=1$ (cross) and 100 (solid). The unfilled
square symbols are for a single source with $\dot{E}=5.8\times
10^{41}$ erg s$^{-1}$. }
\end{figure}

In Figure 9 we plot $d\ln V(t)/d\ln t$ vs. $\ln t$ of the two sources in
Figure 8. Similar to Figure 2, the evolution of $d\ln V(t)/d\ln t$
generally consists of three phases:  the fast growth phase, when $t$
is small, $d\ln V(t)/d\ln t \simeq 3$, the transition phase, and
the slow growth phase, when $t$ is large, $d\ln V(t)/d\ln t$ approaches
to $\sim$ 1. For the ionized region of two sources, one can not
define the I-front with the radius of the ionized region, because the
ionized region is no longer spherical. However, $d\ln V(t)/d\ln t
\simeq 3$ tells us that the length scale of the non-spherical ionized
region should increase with the speed of $\simeq c$.

\begin{figure}
\centering
\includegraphics[width=9cm]{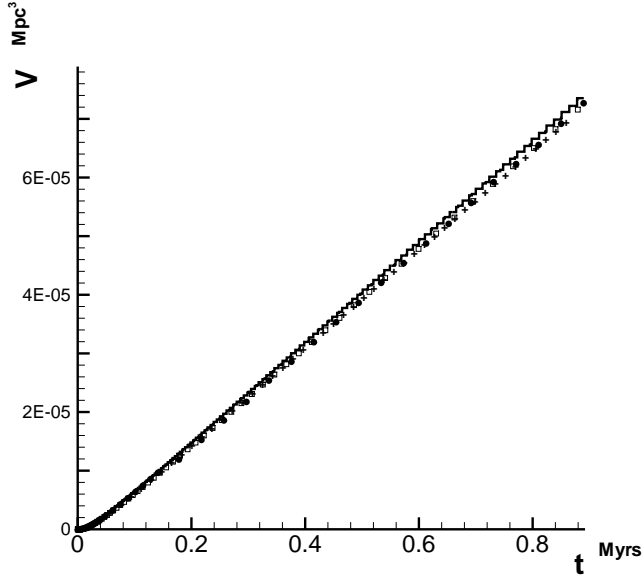}
\caption{
Ionized volume $V(t)$ vs. time $t$ of two sources with the
same parameters as those in Figure 7. $a=100$, 10 and 1 are shown,
respectively, by solid line, filled circle symbols and cross symbols.
The unfilled square symbols are for a single point source with
$\dot{E}=\dot{E}_1+\dot{E}_2$.
}
\end{figure}

As expected, Figure 9 shows that the transition time scale $t_c$ of
the case $a=100$ is shorter than that of the cases $a=1$ and 10. Figure 9
presents also the curve of $d\ln V(t)/d\ln t$ vs. $\ln t$ of the single
point source with $\dot{E}=\dot{E}_1+\dot{E}_2$, which is almost
identical with the curve of $a=1$. Therefore, we can conclude that
in terms of the growth of the ionized volume, two sources with distance $a=1$
and 10 are equal to a single point source with $\dot{E}=\dot{E}_1+\dot{E}_2$.

Figure 10 is similar to Figure 8, but for the two sources with
parameters used in Figure 7. In this case, most UV photons come from
source 1 with $\dot{E}=0.9\times5.8\times 10^{41}$ erg s$^{-1}$, and
the intensity of source 2 $\dot{E}=0.1\times5.8\times 10^{41}$ erg
s$^{-1}$ is much smaller than that of source 1. Nevertheless, we still can
see that the growth of the total ionized volume of $a=100$ is faster
than that of $a=1$ and 10. The growth of the ionized volume of the cases
$a=1$ and 10 are the same, and it is also the same as that of the single
source with intensity $\dot{E}=\dot{E}_1+\dot{E}_2$.

\begin{figure}
\centering
\includegraphics[width=9.0cm, angle=0]{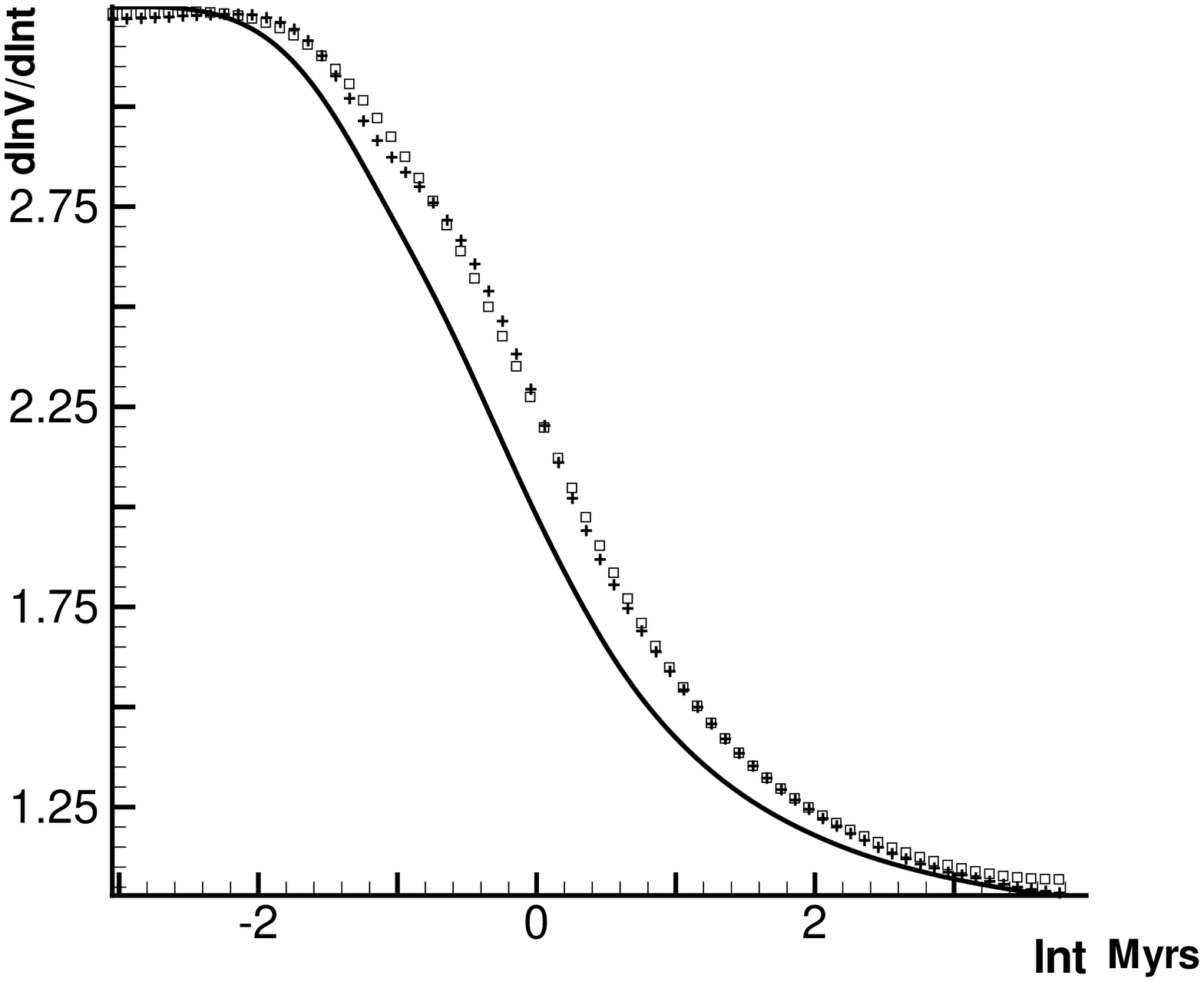}
\caption{ $d\ln V(t)/d\ln t$ vs. $\ln t$ (variable $t$ is
dimensionless) of two sources with intensity
$\dot{E}=0.9\times5.8\times 10^{41}$ and $0.1\times 5.8\times
10^{41}$ erg s$^{-1}$ at $a=1$ (cross) and 100 (solid). The unfilled
square symbols are for a single source with $\dot{E}=5.8\times
10^{41}$ erg s$^{-1}$. }
\end{figure}

Figure 11 is similar to Figure 9, but for the two sources shown in
Figure 10. Similar to Figure 9, the transition time scale $t_c$ of
the case $a=100$ is shorter than that of the cases $a=1$ and 10. The
curve of
$d\ln V(t)/d\ln t$ vs. $\ln t$ for $a=1$ is almost identical to
the curve of a single point source with
$\dot{E}=\dot{E}_1+\dot{E}_2$.

Thus, in terms of the growth of the ionized volume, once the two sources
$\dot{E}_1$ and $\dot{E}_2$ have a distance $<c(t_{1c}+t_{2c})$, one
can replace them by a single source of
$\dot{E}=\dot{E}_1+\dot{E}_2$. Furthermore, one may apply the merging
of two sources to multi-sources. In other words, in a cluster of sources,
we can use the so-called friend-of-friend method to identify all the
sources, for which the distances between all the nearest neighbors $i$
and $j$ are smaller than $c(t^i_c+t^j_c)$, $t^i_c$ being the time scale
$t_c$ of source $i$. That is, for each point source, we plot a
sphere around the source with radius $ct_c$ corresponding to its
intensity, two sources with overlapped spheres can be identified as
a cluster consisting of the two sources. Repeatedly applying the
method to each pair of sources, one can then identify a cluster
consisting of all the sources of which the $ct_c$ spheres are
connected. For this cluster, the ionized volume growth
can be approximately described by a single source of
$\dot{E}=\sum_i\dot{E}_i$, where $\dot{E}_i$ is the intensity
of source $i$.

\section{Discussion and conclusion}

We have developed the WENO algorithm to solve radiative transfer
equation beyond one physical dimension, which can precisely reveal
the features of the merging of the ionized regions of UV photon
sources.

We show that the growth of the ionized volume around either one or two
point sources generally consists of three phases: fast growth phase
at the early evolution, slow growth phase at the later evolution,
and a transition phase between them. The
transition time $t_c$ depends significantly on the intensity of
the photon sources $\dot{E}$, approximately to be
$t_c \propto \dot{E}^{1/2}$.
Most of the photons emitted by the ionizing sources would be
delayed a time $t_c$ to contribute to the ionization. The longer the
$t_c$ the less effective the ionization. Consequently, linear
superposition is not available in estimating the ionized volume. The
growth of the ionized volume of multiple isolated sources $\dot{E}_i$
($i= 1, 2, \dots$)
generally is much faster than that of a single source with intensity
$\dot{E}=\Sigma_i\dot{E}_i$.

Therefore, to calculate accurately the growth of the ionized volume
at $t$, one should not use eq.(\ref{eq1}), but take into account of
the retardation of UV photons. This effect is more substantial if the
first generation of stars are highly clustered. If the ionized spheres
of these sources are merging in relativistic growth phase, the cluster
would behave as a single source with intensity equal to the summation
of intensities of these sources. Therefore, tight clustering of UV
sources will delay the development of reionization.

These results may already be useful in studying the 21 cm signals
from the reionization epoch. It has been shown that a 21 cm emission and
absorption region will develop around a point source once the speed
of the ionization front (I-front) is significantly lower than the
speed of light (Liu et al. 2007). The 21 cm region extends from the
I-front to the front of light $(r=ct)$; its inner part is the
emission region and its outer part is the absorption region. Therefore,
the 21 cm region should be formed only at the time $t>t_c$. Since sources
with weak intensity have small $t_c$, while strong sources or tight
clusters of weak sources have longer $t_c$, sources with weak
intensity produce 21 cm signal earlier, while strong sources, or
tightly clustered weak sources produce it later. The results
calculated from equation (1) will over-predict the ionization,
and thus lead to less 21 cm signals. These features could be
important to reconstruct the history of reionization
with 21 cm tomography and/or cross correlation between redshifted 21
cm signals and emission on other bands from the center of the
sources.

\noindent{\bf Acknowledgments.}

This work is supported in part by the US NSF under the grants
AST-0506734 and AST-0507340. J.R.Liu is supported partially by the
ICRAnet.

\appendix

\section{Equation}

The radiative transfer equation in an expanding universe is
(Bernstein, 1988; Qiu et al. 2006)
\begin{equation}
\label{eqa1}
        {\partial J\over\partial \, (ct)} +
        \frac{1}{a}\nabla \cdot {\bf n}J
        +
        {\partial \over\partial \omega}(HJ) =
        - (k_\nu+3H) J + S,
\end{equation}
where $J(t, {\bf x}, \nu, {\bf n})$ is the specific intensity, $a$
is the cosmic factor, $H=\dot{a}/a$, $\nu$ is the frequency of photon,
$\omega\equiv \ln 1/\nu$, and ${\bf n}$ is a unit vector in the
direction of the photon propagation. We take $c=1$ below. The absorption
coefficient $k_{\nu}$ is
\begin{equation}
\label{eqa2}
k_{\nu} = \sigma(\nu)n_{{\rm HI}}(t, {\bf x})
\end{equation}
where the cross section
$\sigma(\nu)=\sigma_0(\nu_0/\nu )^3$ and $\sigma_0=6.3\times
10^{-18}$  cm$^2$.

Assuming that the point sources $i$ ($i=1, \dots , n$) are located at $x_i$, and
all photons are emitted along the radial direction, we have
\begin{equation}
\label{eqa3}
S(t, {\bf x},{\nu}, {\bf n})=\sum_{i=1}^n f_i(t, {\bf x}, \nu)
   \delta({\bf n - e_{r_i}}),
\end{equation}
where ${\bf e_{r_i}}$ is the unit radial vector with respect to
${\bf x}_i$. The function $f_i$ is given by
\begin{equation}
\label{eqa4}
      f_i(t, {\bf x}, \nu)=\left \{ \begin{array}{ll}
      \dot{E}_i(\nu)/V, & {\rm at \ {\bf x}={\bf x_i}} \\
      0,  & {\rm otherwise},
      \end{array} \right .
\end{equation}
where $\dot{E}_i(\nu)$ is the energy of photons emitted from the sources
$i$ per unit time at frequency $\nu$.
When $V\rightarrow 0$, $f_i(t, {\bf x}, \nu)\rightarrow
\dot{E}_i \delta ({\bf x}_i)$.  We assume the energy spectrum of
UV photons to
be of a power law $\dot{E}(\nu)=\dot{E_0}(\nu_0/\nu)^{\alpha}$, and
$\nu_0$ is the ionization energy of the ground state of hydrogen
$h\nu_0=13.6$ eV. Integration of $\dot{E}$ over $\nu$ gives the
total intensity (energy per unit time) of ionizing photons emitted
by the source, $\dot{E}=\int_{\nu_0}^{\infty} \dot{E}(\nu)d\nu =
\dot{E}_0 \nu_0/(\alpha-1)$.

Since there is no photon-photon collision, the solution of eq.(\ref{eqa1})
can be written as
\begin{equation}
\label{eqa5}
J= \sum_{i=1}^n J_i\delta({\bf n - e_{r_i}})
\end{equation}
and $J_i$ satisfies following equation
\begin{equation}
\label{eqa6}
{\partial J_i\over\partial \, t} +
        \nabla \cdot {\bf e_{r_i}}J_i
         =
        - k_\nu J_i + f_i.
\end{equation}

The coupling among $J_i$ is via the absorption coefficient $k_\nu$
defined in eq.(\ref{eqa2}).  The evolution of the number
density of neutral hydrogen, $n_{\rm HI}(t, {\bf x})$, is
governed by the ionization equation,
\begin{equation}
\label{eqa7}
\frac{df_{\rm HI}}{dt}=\alpha_{\rm HII}n_ef_{\rm HII} - \Gamma_{\rm \gamma HI}f_{\rm HI}- \Gamma_{\rm e HI}n_ef_{\rm HI}
\end{equation}
where the fraction of neutral hydrogen $f_{\rm HI}\equiv n_{\rm HI}(t, {\bf x})/n({\bf x})$, $n({\bf x})$ is the hydrogen distribution,
and $n_e=n({\bf x})-n_{\rm HI}(t, {\bf x})$ is the number density of electrons, if the electrons from ionized helium can be ignored.
$\alpha_{\rm HII}$ is the recombination coefficient,
$\Gamma_{\rm eHI}$ is the collision ionization rate,
and the photoionization rate $\Gamma_{\rm \gamma HI}(t, \rho, z)$ is given by
\begin{equation}
\label{eqa8}
\Gamma_{\gamma {\rm HI}}(t,{\bf x})=\sum_{i=1}^n
\int_{\nu_0}^{\infty} d\nu \frac{J_{i}(t,{\bf x}, \nu)}{h\nu}\sigma(\nu).
\end{equation}

The evolution of the hydrogen gas temperature $T$, in the unit of K, is given by
\begin{equation}
\label{eqa9}
\frac{dU}{dt}=H-n^2C,
\end{equation}
where $U=\frac{3}{2}n k_BT $ with $k_B$ be the Boltzmann constant, and
\begin{equation}
\label{eqa10}
H = n_{\rm HI}\sum_{i=1}^n
\int_{\nu_0}^\infty J_i(t, {\bf x}, \nu)
\sigma(\nu)\frac{\nu-\nu_0}{\nu}d\nu.
\end{equation}

The relevant parameters are as follows (Theuns et al. 1998)

1. The recombination coefficient
\begin{equation}
\label{eqa11}
\alpha_{\rm HII}=
 6.30\times10^{-11}T^{-1/2}T_3^{-0.2}/(1+T^{0.7}_6),
\end{equation}
where $T$ is temperature, and $T_n=T/10^n$.

2. The collision ionization
\begin{equation}
\label{eqa12}
\Gamma_{\rm eHI}=
    1.17\times 10^{-10}T^{1/2}e^{-157809.1/T}(1+T_5^{1/2})^{-1}.
\end{equation}

3. The cooling. Since only the recombination cooling is important, we have
\begin{eqnarray}
\label{eqa13}
C  & = & 8.70\times 10^{-27}T^{1/2}T_3^{-0.2}
   (1+T_6^{0.7})^{-1}[1-f_{\rm HI}]^2 \\ \nonumber
   & + & 1.42\times 10^{-27} T^{1/2}[1-f_{\rm HI}]^2 \\ \nonumber
   & + & 2.45\times 10^{-21}T^{1/2} e^{-157809.1/T}(1+T_5^{1/2})^{-1}
   (1-f_{\rm HI})f_{\rm HI}
     \\\nonumber
   & + & 7.5\times 10^{-19} e^{-118348/T}(1+T_5^{1/2})^{-1}
(1-f_{\rm HI})f_{\rm HI}
\end{eqnarray}
where $T_n=T/10^n$. The terms on the r.h.s. of eq.(\ref{eqa13}) are,
respectively, the recombination cooling, the collisional ionization
cooling, the collisional excitation cooling and bremsstrahlung.
Both $H$ and $C$ are in the unit of ergs cm$^{3}$ s$^{-1}$.

\section{Two sources}

Consider the case of two sources.
Using cylindrical coordinate $(\rho, \theta, z)$,
the two sources are assumed to be located at ${\bf x}={\bf x}_{\pm}=(0, 0, a)$, with
${\bf e}_{r_{\pm}}=(\rho/\sqrt {\rho^2+(z\mp a)^2}, 0, (z\mp a)/\sqrt {\rho^2+(z\mp a)^2})$.
The corresponding radiative transfer equations eq.(\ref{eqa6}) are
\begin{eqnarray}
\label{qiuadd}
\lefteqn { {\partial J_{\pm}\over\partial \, t} +
  \frac{1}{\rho}\frac{\partial}{\partial \rho}\left (\frac{\rho^2}
   {\sqrt {\rho^2+(z \mp a)^2}}J_{\pm}\right ) +
   \frac{\partial}{\partial z}\left ( \frac{(z \mp a)}
   {\sqrt {\rho^2+(z \mp a)^2}}J_{\pm} \right ) } \\ \nonumber
  & & \hspace{3cm}= - k_\nu J_{\pm} + f_{\pm},
\end{eqnarray}
where $J_{\pm}(t, \rho, z, \nu)$ are the specific intensities for sources ${\bf x_{\pm}}$.

Instead of adding the source term $f_{\pm}$ in the r.h.s of eq.(\ref{qiuadd}), equivalently,
we impose a boundary condition
\begin{equation}
\label{eqa14}
\lim_{\rho\rightarrow 0, z \rightarrow \pm a}
4\pi r_{\pm}^2 J_{\pm}(t,\rho, z, \nu)=\dot{ E}_{\pm}(\nu),
\end{equation}
where $r_{\pm}=\sqrt{\rho^2 + (z\mp a)^2}$,
$\dot{E}(\nu)=\dot{E_0}_{\pm}(\nu_0/\nu)^{\alpha}$ is the energy of photons emitted from the sources per unit time at frequency $\nu$.

The absorption coefficient in eq.(\ref{qiuadd}) $k_{\nu}$ is defined in eq.(\ref{eqa2}),
with $n_{\rm HI}(t, \rho, z)\equiv n f_{\rm HI}(t, \rho, z)$ governed by the ionization equation eq.(\ref{eqa7}).
Here the photoionization rate $\Gamma_{\rm \gamma HI}(t, \rho, z)$ in eq.(\ref{eqa7}) is given by
\begin{equation}
\label{eqa16}
 \Gamma_{\rm \gamma HI}(t, \rho, z)=
\int_{\nu_0}^{\infty} d\nu \frac{J_{+}(t,\rho, z, \nu)+J_{+}(t,\rho, z, \nu)}{h\nu}\sigma(\nu).
\end{equation}
The kinetic temperature of the baryon gas is determined by eq.(\ref{eqa9}) with
\begin{equation}
\label{qiuadd2}
H = n_{\rm HI} \int_{\nu_0}^\infty (J_+(t, \rho, z, \nu)+ J_-(t,\rho,z, \nu))\sigma(\nu)\frac{\nu-\nu_0}{\nu}d\nu.
\end{equation}

\section{The numerical algorithm}

In this paper, we are solving the system of equations (\ref{qiuadd}), (\ref{eqa7}) and (\ref{eqa9}).
To apply the WENO algorithm, we rewrite eq.(\ref{qiuadd}) into a conservative form as
\begin{equation}
\label{eqC0}
\frac{\partial J_{\pm}}{\partial t} +
\frac{\partial}{\partial \rho}[\frac{\rho}{r_{\pm}}J_{\pm}] +
\frac{\partial}{\partial z}[\frac{z \mp a}{r_\pm}J_{\pm}] = -
\frac{1}{r_{\pm}} J_{\pm} - k_\nu J_{\pm},
\end{equation}
where $r_{\pm} = \sqrt{\rho^2 + (z\mp a)^2}$.
We use the WENO algorithm to approximate the spatial derivatives in eq.(\ref{eqC0}) (Qiu et al. 2006).

In the numerical implementation, it is convenient to introduce the
dimensionless variables $t'$, $\rho'$, $z'$, $a'$, $\nu'$, $J'$ by
rescaling $t' = c n \sigma_0 t$, $\rho' = n \sigma_0 \rho$, $z' = n
\sigma_0 z$, $a' = n \sigma_0 a$, $\nu'=\nu/\nu_0$, and $J'_{\pm}
d\nu' = \frac{\sigma_0}{h \nu_0 n}J_{\pm} d\nu$.
Therefore, $t'$ and $r'$ are respectively, the time and distance in the
units of mean free flight
time and mean free path of ionizing photon $h\nu_0$ in the
non-ionized background hydrogen gas $n$. For the $\Lambda$CDM model,
$n=1.88\times 10^{-7}(1+z)^3$ cm$^{-3}$, where $z$ is the redshift,
$t=0.89 (1+z)^{-3} t'$ Myrs and $r=0.27(1+z)^{-3}r'$ Mpc.
Then the system of equations (\ref{eqC0}), (\ref{eqa7}) and (\ref{eqa9}) can be
rewritten as the following system
\begin{equation}
\label{eqC1}
\frac{\partial J'_{\pm}}{\partial t'} +
\frac{\partial}{\partial \rho'}[\frac{\rho'}{r'_{\pm}}J'_{\pm}] +
\frac{\partial}{\partial z'}[\frac{z'\mp a'}{r'_\pm}J'_{\pm}] = -
\frac{1}{r'_{\pm}} J'_{\pm} - \frac {1}{\nu'^3}f_{HI} J'_{\pm}
\end{equation}
\begin{equation}
\label{eqC2} c \sigma_0 \frac{d f_{HI}}{dt'} = \alpha_{HII} (1-
f_{HI})^2 - \frac{\Gamma_{\gamma HI}}{n} f_{HI} - \Gamma_{eHI} (1-
f_{HI}) f_{HI}
\end{equation}
\begin{equation}
\label{eqC3} \frac 32 c \sigma_0 k_B \frac{\partial T}{\partial t'}
= H-C
\end{equation}
with $\frac{\Gamma_{\gamma HI}}{n}$ given by
\begin{equation}
\frac 1n \Gamma_{\gamma HI}(t, \rho, z) = \int_1^\infty  \frac{J'_+
+ J'_{-}}{\nu'^4} d \nu',
\end{equation}
$H$ given by
\begin{equation}
H = h\nu_0 f_{HI} \int_1^\infty (J'_+ + J'_-) \frac{\nu'-1}{\nu'^4}
d\nu',
\end{equation}
and $\alpha_{HII}$, $\Gamma_{eHI}$ and $C$ given by equations (\ref{eqa11}),
(\ref{eqa12}) and (\ref{eqa13}) respectively.

To solve the radiative transfer equation (\ref{eqC1}), we adopt
the fifth-order finite difference WENO scheme, which was designed in
(Jiang $\&$ Shu 1996), coupled with the third order TVD Runge-Kutta time
discretization for the system of equations (\ref{eqC1}), (\ref{eqC2})
and (\ref{eqC3}).  The multi-time-scale strategy (Qiu et al. 2007) and
the adaptive time step strategy (Liu et al. 2007) are used to save the
increased computational cost introduced by the stiffness of the equations
(\ref{eqC2}) and (\ref{eqC3}). The numerical algorithms are implemented
as describe below. For the sake of simplicity, we drop the prime in the
notations. For example, $J$ means $J'$ hereafter.

\begin{itemize}

\item
The computational domain and computational mesh:

The computational domain is $(\rho, z, \nu) \in [0, \rho_{max}]
\times [-z_{max}, z_{max}] \times [1, \nu_{max}]$, where
$\rho_{max}$ and $z_{max}$ are chosen such that $J(t, \rho, z,
\nu)\approx 0$ for $\rho >  \rho_{max}$, $|z| > z_{max}$ or $\nu >
\nu_{max}$. In our computation, $\rho_{max}$ is taken to be greater
than the final computational time t, $z_{max}$ is taken to be
$\rho_{max} + a$ and $\nu_{max}=10^6$.

To avoid $r_{\pm}$ in eq. (\ref{eqC1}) to become 0, we
design the computational mesh, such that ${\bf x_{\pm}}$ are located
at the center between grid points. The mesh sizes in the $\rho$- and $z$- directions are
set to be the same, i.e. $\Delta \rho = \Delta z$, to preserve the
spherical symmetry property of the numerical solution before the merging of
the two sources.
The mesh in the $\nu$ direction is taken to be smooth but not uniform.
Specifically, the mesh sizes are designed as the following
$$
\Delta z = \frac{a}{N_{z_a} + \frac12}; \quad \Delta \rho =
\Delta z; \quad \Delta \xi = \log_2 {\nu_{max}}/N_\nu;
$$
with $N_{z_a}$ being the number of mesh points in [0, a] in
the $z$-direction, and $N_\nu$ being the number of mesh points in
the $\nu$-direction.  The computational mesh is
$$
\rho_i = \left( i - \frac12 \right) \Delta \rho, \quad i = 1, 2, ..., N_\rho,
$$
$$
z_j = j \Delta z, \quad j = 1, 2, ..., N_z,
$$
$$
\nu_k = 2^{\xi_k}, \quad with \quad \xi_k = k \Delta \xi, \quad k = 1, 2, ..., N_\nu .
$$

\item
The WENO method in approximating the spatial derivatives:

The approximation to the point values of the solution $J_{\pm} (t^n,
\rho_i, z_j, \nu_k)$, denoted by $J^n_{\pm, i, j, k}$, is obtained
with a dimension by dimension approximation to the spatial
derivatives using the fifth order WENO scheme (Jiang $\&$ Shu 1996).
Taking $\frac{\partial}{\partial z} (\frac{z - a}{r_{+}} J_{+})$ as
an example, the approximation is performed along the $z$-line with fixed
$\rho_i$ and $\nu_k$:
\begin{equation}
\frac{\partial}{\partial z}  \left( \frac{z_j - a}{r_{+} (\rho_i, z_j)}
J_{+}(t^n, \rho_i, z_j, \nu_k) \right) \approx \frac{1} {\Delta z}
(\hat{h}_{j+1/2}-\hat{h}_{j-1/2})
\end{equation}
where the numerical flux $\hat{h}_{j+\frac12}$ is obtained with the
following procedure. When the ``wind direction'', namely the
coefficient $\frac{z_{j} + z_{j+1}}{2} - a$ is positive at the mesh
boundary, we use a left-biased stencil in reconstructing the
numerical flux $\hat{h}_{j+\frac12}$ as described in detail below.
When the ``wind direction'' is negative, we use a right-biased stencil
to obtain the numerical flux $\hat{h}_{j+\frac12}$, following a
mirror symmetry reconstruction with respect to $j+\frac12$
as that of the left-biased stencil. When the coefficient
$\frac{z_{j} + z_{j+1}}{2} - a$ = 0, which actually will happen due
to the way we design our mesh, the numerical flux is simply set to
be 0.

\noindent We denote
$$
{h_j} = J(t^n, \rho_i, z_j, \nu_k), \qquad j=-2, -1, ..., {N_z}+2
$$
where $n$, $i$ and $k$ are fixed. The numerical flux from the
regular WENO procedure is obtained by
$$
\hat{h}_{j+1/2} = \omega_1 \hat{h}_{j+1/2}^{(1)} + \omega_2
\hat{h}_{j+1/2}^{(2)} + \omega_3 \hat{h}_{j+1/2}^{(3)}
$$
where $\hat{h}_{j+1/2}^{(m)}$ are the three third order fluxes on
three different stencils given by
\begin{eqnarray*}
\hat{h}_{j+1/2}^{(1)} & = & \frac{1}{3} h_{j-2} - \frac{7}{6}
h_{j-1}
               + \frac{11}{6} h_{j}, \\
\hat{h}_{j+1/2}^{(2)} & = & -\frac{1}{6} h_{j-1} + \frac{5}{6} h_{j}
               + \frac{1}{3} h_{j+1}, \\
\hat{h}_{j+1/2}^{(3)} & = & \frac{1}{3} h_{j} + \frac{5}{6} h_{j+1}
               - \frac{1}{6} h_{j+2},
\end{eqnarray*}
and the nonlinear weights $\omega_m$ are given by
$$
\omega_m = \frac {\tilde{\omega}_m} {\sum_{l=1}^3
\tilde{\omega}_l},\qquad
 \tilde{\omega}_l = \frac {\gamma_l}{(\varepsilon + \beta_l)^2} ,
$$
with the linear weights $\gamma_l$ given by
$$
\gamma_1=\frac{1}{10}, \qquad \gamma_2=\frac{3}{5}, \qquad
\gamma_3=\frac{3}{10},
$$
and the smoothness indicators $\beta_l$ given by
\begin{eqnarray*}
\beta_1 & = & \frac{13}{12} \left( h_{j-2} - 2 h_{j-1}
                             + h_{j} \right)^2 +
         \frac{1}{4} \left( h_{j-2} - 4 h_{j-1}
                             + 3 h_{j} \right)^2  \\
\beta_2 & = & \frac{13}{12} \left( h_{j-1} - 2 h_{j}
                             + h_{j+1} \right)^2 +
         \frac{1}{4} \left( h_{j-1}
                             -  h_{j+1} \right)^2  \\
\beta_3 & = & \frac{13}{12} \left( h_{j} - 2 h_{j+1}
                             + h_{j+2} \right)^2 +
         \frac{1}{4} \left( 3 h_{j} - 4 h_{j+1}
                             + h_{j+2} \right)^2 .
\end{eqnarray*}
$\varepsilon$ is a parameter to avoid the denominator to become 0
and is taken as $\varepsilon = 10^{-5}$ times the maximum magnitude
of the initial condition $J$ in the computation.

\item Time integration:

To evolve in time, we use the third order TVD Runge-Kutta method (Shu $\&$
Osher 1988). For systems of ODEs ${u}_t = L({u})$, the third order
Runge-Kutta method is
\begin{eqnarray}
\label{rk1}
u^{(1)} & = & u^n + \Delta t L(u^n, t^n)  \\
\label{rk2}u^{(2)} & = & \frac 3 4 u^n + \frac 1 4 (u^{(1)} + \Delta t L(u^{(1)})) \\
\label{rk3}u^{n+1} & = & \frac 13 u^n + \frac 23 (u^{(2)}+\Delta t
L(u^{(2)}))
\end{eqnarray}
The difficulty of the direct implementation of the Runge-Kutta
method lies in the stiffness of equations (\ref{eqC2}) and (\ref{eqC3}).
Especially for the strong source, one needs a very small time step $\Delta
t$, as small as $10^{-7}$, to guarantee the stability of the
numerical scheme, therefore the computational cost for long time
integration is huge. In this paper, we adapt the multi-time-scale strategy (Qiu et
al. 2007) and the adaptive-time-step strategy (Liu et al. 2007) to
evolve the system of equations in time. We refer to (Qiu et al.
2007) and (Liu et al. 2007) for the details of its implementation.

\item Numerical boundary condition:

\begin{itemize}

\item  In the $\rho$-direction,\\
at $\rho =0$,
$$
J_{\pm, -i, j, k} = J_{\pm, i+1, j, k}, \quad i=0, 1, 2,
$$
at $\rho = \rho_{max}$,
$$
J_{\pm, N_\rho+i, j, k} = 0, \quad i = 0, 1, 2.
$$
\item In the $z$-direction, \\
at $z= \mp z_{max}$,
$$
J_{\pm, i, \mp(N_z+i), k} = 0, \quad i=0, 1, 2
$$

\item
Around the point sources ${\bf x_\pm}$, according to eq.(\ref{eqa14}), \\
$$
J_{\pm, i, j, k} = \frac{1}{\nu_k^\alpha}\frac{\dot E}{ 4 \pi r^2_{\pm, i, j}},
\quad r_{\pm, i, j} < r_s
$$
with $ r_{\pm, i, j} = \sqrt {\rho_i^2 + (z_j\mp a)^2}$. $r_s$ is a
small number depending on the mesh size. $r_s$ is bigger when the
mesh is coarser.

\end{itemize}

\item Parallel computing:

The computational cost in solving the system of equations
(\ref{eqC1}), (\ref{eqC2}) and (\ref{eqC3}) is quite large for this
3-dimensional (2-D in the physical space and 1-D in the frequency space)
time dependent problem.
Parallel computing with mpif77 is used to speed up the computation.

\end{itemize}

\end{document}